\documentclass[aps,prb,twocolumn,groupedaddress]{revtex4-2}

\usepackage[T1]{fontenc}
\usepackage{amsmath}
\usepackage{amssymb}
\usepackage{amsfonts}

\usepackage{bbm}
\usepackage{xspace}
\usepackage{color}
\usepackage{ulem} 
\usepackage{placeins}
\usepackage{bm}
\usepackage{siunitx}
\usepackage{nccmath}
\usepackage{mathtools} 
\usepackage{empheq}
\usepackage{enumerate}
\usepackage{graphicx}
\usepackage{dcolumn}
\usepackage[caption=false]{subfig}

\usepackage[colorlinks=true,linkcolor=blue,citecolor=blue,urlcolor=blue]{hyperref}

\usepackage{times}
\usepackage{MnSymbol}

\usepackage[version=4]{mhchem}

\graphicspath{{.}{Bilder/}}

\begin{document}

\title{Pure crossed Andreev reflection assisted transverse valley currents in $\alpha-\mathcal{T}_3$ lattices}
\author{W.~Zeng$^1$}
\author{R.~Shen$^{1,2}$}
\email[E-mail: ]{shen@nju.edu.cn}
\affiliation{$^1$National Laboratory of Solid State Microstructures and School of Physics, Nanjing University, Nanjing 210093, China\\
$^2$Collaborative Innovation Center of Advanced Microstructures, Nanjing University, Nanjing, 210093, China}

%\date{\today}

\begin{abstract}{}
We propose a novel method for the generation of the transverse valley currents, which is based on the pure crossed Andreev reflection (pCAR) in the superconducting hybrid junctions composed of the gapped $\alpha-\mathcal{T}_3$ lattices with ferromagnet-induced exchange interaction. The angle-resolved pCAR probability is asymmetric for a given valley, resulting in the transverse valley currents with zero net charge. This pCAR assisted charge-valley conversion is highly efficient with the valley Hall angle reaching an order of unity, suggesting potential applications for valleytronic devices.
\end{abstract}

\maketitle
\section{Introduction}\label{intro}

The $\alpha-\mathcal{T}_3$ lattice is an extension of the graphene honeycomb lattice with an additional site centered at each hexagonal cell~\cite{PhysRevB.99.045420,PhysRevB.99.205429,PhysRevB.95.235432,PhysRevB.103.165429,PhysRevB.98.075422}. The coupling strength between the additional site and one of the honeycomb subsites is parameterized by $\alpha$, which varies from $\alpha=0$ (graphene lattice) to $\alpha=1$ (dice lattice). The low energy excitations in $\alpha-\mathcal{T}_3$ lattices are the massless pseudospin-one Dirac fermions, which are featured by a flat band cutting through two linearly dispersing branches at the nonequivalent Dirac points $K$ and $K'$~\cite{PhysRevA.96.033634,PhysRevB.100.085134}. A symmetry-breaking term introduces an additional effective mass in the $\alpha-\mathcal{T}_3$ lattice~\cite{PhysRevB.105.165402,Zeng_2022}, leading to the bandgap opening at the Dirac point~\cite{PhysRevB.103.165429}. Several methods have been proposed to realize the $\alpha-\mathcal{T}_3$ lattice in experiments. The dice lattice with $\alpha=1$ can be produced in $\ce{SrTiO3}$/$\ce{SrIrO3}$/$\ce{SrTiO3}$ trilayer heterostructure grown along the ($111$) direction~\cite{PhysRevB.84.241103}. The Hamiltonian of $\ce{Hg_{1-x}Cd_{x}Te}$ at the critical doping can be mapped to that of the $\alpha-\mathcal{T}_3$ lattice with $\alpha=1/\sqrt{3}$~\cite{PhysRevB.92.035118}. Some novel properties of the $\alpha-\mathcal{T}_3$ lattice are attributed to the dispersionless flat band. Such as the super Klein tunneling~\cite{PhysRevB.95.235432,PhysRevResearch.2.043245,PhysRevB.101.035129}, the super Andreev reflection~\cite{Zeng_2022,PhysRevB.104.125441,PhysRevB.101.235417}, the flat-band ferromagnetism~\cite{PhysRevLett.62.1201,PhysRevLett.69.1608}, and the unconventional Anderson localization~\cite{PhysRevLett.113.236403,PhysRevB.82.104209}. 

\begin{figure}[!tp]
\centerline{\includegraphics[width=1\linewidth]{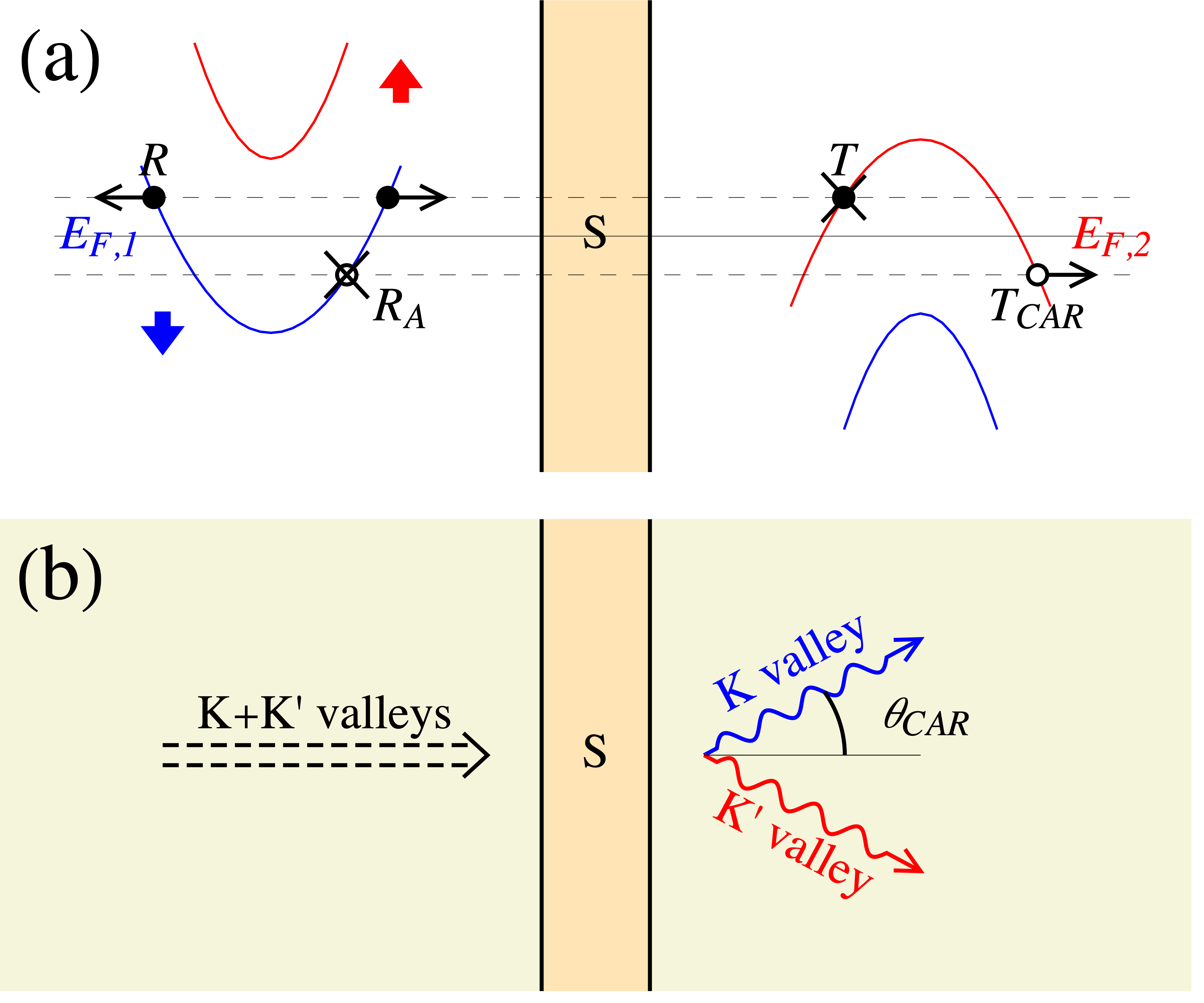}}
\caption{\label{fig:1}
(a) Scattering processes in the spin-split parabolic bands. The spin-up and the spin-down subbands are denoted by the red and blue lines, respectively. The dispersionless flat bands are not shown. The incident electron is local reflected as an electron ($R$) and nonlocal transmitted as a hole ($T_{CAR}$). The local Andreev reflection ($R_A$) and the electron elastic cotunneling ($T$) are blocked due to the spin mismatch. (b) Schematic of the valley-contrasting pCAR in the FSF junctions. The wavy lines denote the crossed Andreev reflected holes.
}
\end{figure}

In addition, the $\alpha-\mathcal{T}_3$ lattice is also a promising candidate for valleytronics~\cite{PhysRevB.96.245410,PhysRevB.101.245416,PhysRevB.82.165409,PhysRevB.103.184403,PhysRevB.87.155415,PhysRevB.105.094510}, where the generation of the controllable valley polarization and valley current is the key issue. Up to now, many strategies have been proposed to do so, such as the valley-polarized current produced in the $\alpha-\mathcal{T}_3$ lattice-based magnetic Fabry-P\'erot interferometer~\cite{PhysRevB.102.075443}, the valley-polarized magnetoconductivity in the periodically modulated $\alpha-\mathcal{T}_3$ lattice~\cite{PhysRevB.96.045418} and the valley filtering in strain-induced $\alpha-\mathcal{T}_3$ quantum dots~\cite{PhysRevB.103.165114}. Analogous to the spin Hall effect~\cite{PhysRevLett.83.1834,PhysRevLett.96.106802,PhysRevLett.92.126603}, the valley Hall effect~\cite{PhysRevLett.99.236809,tong2016concepts,lee2016electrical} is also an alternative solution to generate the controllable valley current. Recently, the geometric valley Hall effect was reported in the $\alpha-\mathcal{T}_3$ lattice~\cite{PhysRevB.96.045412}, where the valley-contrasting scattering and the transverse valley current can be produced by the isotropic valley-independent impurities. Inspired by this, we propose a new method to generate the transverse valley current in the $\alpha-\mathcal{T}_3$ lattice by means of the crossed Andreev reflection (CAR)~\cite{PhysRevLett.93.197003,PhysRevB.78.235403,PhysRevLett.103.237001,PhysRevB.76.224506,PhysRevLett.127.017701}. We focus on the pure crossed Andreev reflection (pCAR) dominated transport~\cite{PhysRevLett.122.257701,PhysRevLett.127.076601,PhysRevB.95.245433,PhysRevB.93.041422} in the ferromagnet/superconductor/ferromagnet (FSF) junctions based on the gapped $\alpha-\mathcal{T}_3$ lattices, where the Fermi level is located between two spin-subband edges, as shown in Fig.~\ref{fig:1}(a). The local Andreev reflection and the electron elastic cotunneling are completely inhibited due to the spin mismatch. For $\alpha\neq0$ and $\alpha\neq1$, the angle-dependence of the CAR probability is asymmetric for a given valley. The crossed Andreev reflected holes in different valleys turn into opposite directions, as shown in Fig.~\ref{fig:1}(b), leading to a transverse valley current. The total CAR probability ($K$ valley $+$ $K'$ valley) is symmetric due to the time-reversal symmetry. Consequently, the transverse charge current is zero. The transverse valley conductance as well as the longitudinal charge conductance is determined by the pCAR process, which can be electrically controlled by tuning the Fermi level and the incident energy. The charge-valley conversions are highly efficient with the valley Hall angle reaching an order of unity, suggesting their great potential for valleytronic applications.

The remainder of the paper is organized as follows. The model Hamiltonian and the scattering approach are explained in detail in Sec.~\ref{sec:2}. The numerical results and discussions are presented in Sec.~\ref{sec:3}. Finally, we conclude in Sec.~\ref{sec:4}.

\section{Model}\label{sec:2}

We consider the FSF junction in the $x-y$ plane with the superconducting electrode covering the region $0<x<d$. The low energy Hamiltonian of the ferromagnetic $\alpha-\mathcal{T}_3$ lattice is given by~\cite{PhysRevB.99.205429}
\begin{align}
\mathcal{H}_{\eta\sigma}(\bm k)=
&\begin{bmatrix}
0 &f_\eta(\bm k)\cos\varphi  &0 \\ 
f^*_\eta(\bm k)\cos\varphi&0  &f_\eta(\bm k)\sin\varphi \\ 
0&f^*_\eta(\bm k)\sin\varphi  &0 
\end{bmatrix}\label{eq:ht}\\
\nonumber&+\Delta U+\sigma h,
\end{align}
where $f_\eta(\bm k)=\hbar v(\eta k_x-ik_y)$, $k_{x(y)}$ is the wave vector in the $x$ ($y$) direction, $v$ is the Fermi velocity, $\eta=+$ $(-)$ for $K$ ($K'$) valley, $\sigma=+$ $(-)$ for the spin-up (spin-down) electrons, $\Delta$ measures the sublattice symmetry breaking with the corresponding matrix $U=\mathrm{diag}[1,-1,1]$, resulting in the massive Dirac fermions with the effective mass $m=\Delta/v^2$~\cite{PhysRevB.105.165402}. The parameter $\varphi=\tan^{-1}\alpha$ provides a continuous lattice transformation from the graphene-like lattice ($\alpha=0$) to the dice lattice ($\alpha=1$). The ferromagnetic exchange energy $h$ is only applied in the normal region, which can be induced by the proximity to an insulating ferromagnetic layer~\cite{PhysRevB.100.245148,PhysRevB.102.075443}. The energy dispersion can be directly obtained from Eq.~(\ref{eq:ht}). The valley-degenerate parabolic bands are given by
\begin{align}
    E_{\lambda\sigma}=\lambda\sqrt{(\hbar v\bm k)^2+\Delta^2}-E_F+\sigma h,
\end{align}
where $\lambda=+$ $(-)$ for the conduction (valence) bands. The dispersionless flat bands are located at the bottom of the conduction bands, which are given by $E_{0,\sigma}=-E_F+\Delta+\sigma h$.

The Dirac-Bogoliubov-de Gennes (DBdG) equation describing the quasiparticle excitations in the superconducting region reads~\cite{PhysRevLett.97.067007,de2018superconductivity}
\begin{align}
\begin{bmatrix}
\mathcal{H}-E_F &\Delta_{sc}(x) \\ 
\Delta^*_{sc}(x) & E_F-\mathcal{T}\mathcal{H}\mathcal{T}^{-1}
\end{bmatrix}
\begin{bmatrix}
u\\ 
v
\end{bmatrix}=E\begin{bmatrix}
u\\ 
v
\end{bmatrix},\label{bdg}
\end{align}
where $\mathcal{H}=\mathrm{diag}[\mathcal{H}_{+\uparrow},\mathcal{H}_{+\downarrow},\mathcal{H}_{-\uparrow},\mathcal{H}_{-\downarrow}]$ is the $12\times12$ electron Hamiltonian spanned by the valley, spin and sublattice space, the vector $u$ $(v)$ is the electron (hole) component of the quasiparticle wave function, the excitation energy $E$ is measured from the Fermi level $E_F$. $\mathcal{T}=i\tau_x\sigma_yS_0\mathcal{C}$ is the time-reversal operator, where $\tau_x$ and $\sigma_y$ are the Pauli matrix in the valley and spin space, respectively, $S_0$ is the identity matrix in the sublattice space, and the operator $\mathcal{C}$ denotes the complex conjugation. The $s$-wave superconducting pair potential $\Delta_{sc}(x)$ is zero in the normal region and is $\Delta_{sc}$ in the superconducting region, which can be generated via the proximity effect~\cite{PhysRevB.104.125441,Zeng_2022,PhysRevB.101.235417}. Eq.~(\ref{bdg}) can be decoupled into $4$ subsets due to the fact that the $s$-wave Cooper pairs are composed of the spin-up (down) electrons in the $K$ valley and the spin-down (up) electrons in the $K'$ valley.

For the convenience of calculation, we perform a $\pi/2$ rotation of the crystal coordinates, leading to the combination of the valley index $\eta$ and the conserved transverse wave vector $k_y$. The explicit form of the decoupled DBdG equation with the valley and spin index $(\eta,\sigma)$ reads

\begin{widetext}
\begin{align}
\begin{bmatrix}
\Delta+\sigma h-E_F &\hbar vk^\eta_-\cos\varphi  &0  & \sigma\Delta_{sc}(x) &0  &0 \\ 
\hbar vk^\eta_+\cos\varphi & -\Delta+\sigma h-E_F & \hbar vk^\eta_-\sin\varphi &0  &\sigma\Delta_{sc}(x)  &0 \\ 
0 &\hbar vk^\eta_+\sin\varphi  &\Delta+\sigma h-E_F  &0  &0  &\sigma\Delta_{sc}(x) \\ 
\sigma\Delta_{sc}^*(x) & 0 & 0 & -\Delta+\sigma h+E_F  & -\hbar vk^\eta_-\cos\varphi &0 \\ 
0 &\sigma\Delta_{sc}^*(x)  &0  & -\hbar vk^\eta_+\cos\varphi &\Delta+\sigma h+E_F  &-\hbar vk^\eta_-\sin\varphi \\ 
0 &0  &\sigma\Delta_{sc}^*(x)  &0  &-\hbar vk^\eta_+\sin\varphi  & -\Delta+\sigma h+E_F
\end{bmatrix}\begin{bmatrix}
u^A_\sigma\\ 
u^B_\sigma\\ 
u^C_\sigma\\ 
v^A_{\bar{\sigma}}\\ 
v^B_{\bar{\sigma}}\\ 
v^C_{\bar{\sigma}}
\end{bmatrix}=E\begin{bmatrix}
u^A_\sigma\\ 
u^B_\sigma\\ 
u^C_\sigma\\ 
v^A_{\bar{\sigma}}\\ 
v^B_{\bar{\sigma}}\\ 
v^C_{\bar{\sigma}}
\end{bmatrix},\label{eq:dbdg}
\end{align}
\end{widetext}
where $k^\eta_\pm=k_x\pm i\eta k_y$, $\bar{\sigma}\equiv-\sigma$, and the vector $[u^A_{\sigma}\hdots v^C_{\bar{\sigma}}]^T$ is the eigenstate describing the quasi-particle excitations. In the normal region with $\Delta_{sc}(x)=0$, the scattering wave functions for the electron states are given by 
\begin{align}
\nonumber\psi_{\lambda\eta\sigma}^{(e)\pm}=&\left(\frac{E+E_F-\Delta-\sigma h}{E+E_F-\sigma h}\right)^{\frac{1}{2}}\begin{bmatrix}
\frac{\hbar v(\lambda k_x-i\eta k_y)\cos\varphi}{E+E_F-\Delta-\sigma h}\\ 
1\\ 
\frac{\hbar v(\lambda k_x+i\eta k_y)\sin\varphi}{E+E_F-\Delta-\sigma h}\\ 
0\\ 
0\\ 
0
\end{bmatrix}\\
&\times \exp(ik_xx+ik_yy),
\end{align}
where $k_x=\sqrt{(E+E_F-\sigma h)^2-\Delta^2-(\hbar vk_y)^2}/\hbar v$ is the longitudinal wave vector and the superscript `$\pm$' of the wave function denotes the direction of propagation. The scattering wave functions for the hole states are given by
\begin{align}
\nonumber\psi^{(h)\pm}_{\lambda'\bar{\eta}\bar{\sigma}}=&\left(\frac{E-E_F+\Delta+\bar{\sigma} h}{E-E_F+\bar{\sigma} h}\right)^{\frac{1}{2}}\begin{bmatrix}
0\\
0\\
0\\
\frac{\hbar v(\lambda' k'_x+i\bar{\eta}k_y)\cos\varphi}{E-E_F+\Delta+\bar{\sigma} h}\\ 
-1\\ 
\frac{\hbar v(\lambda' k'_x-i\bar{\eta}k_y)\sin\varphi}{E-E_F+\Delta+\bar{\sigma} h}\\ 
\end{bmatrix}\\
&\times \exp(ik'_xx+ik_yy),
\end{align}
where the longitudinal wave vector for the hole states is $k'_x=\sqrt{(E-E_F-\sigma h)^2-\Delta^2-(\hbar vk_y)^2}/\hbar v$, $\bar{\eta}\equiv-\eta$, and $\lambda'=+$ $(-)$ denotes the conduction (valence) bands of the hole excitations.

In the superconducting region with $h=0$, the scattering states are given by
\begin{align}
\nonumber\psi^{(s)}_{\varsigma\varrho}=&\begin{bmatrix}
\frac{\hbar v(\varsigma k_s-ik_y)}{\mu_s}\cos\varphi e^{i\varrho\frac{\beta}{2}}\\
e^{i\varrho\frac{\alpha}{2}}\\
\frac{\hbar v(\varsigma k_s+ik_y)}{\mu_s}\sin\varphi e^{i\varrho\frac{\beta}{2}}\\
\frac{\hbar v(\varsigma k_s-ik_y)}{\mu_s}\cos\varphi e^{-i\varrho\frac{\beta}{2}}\\ 
e^{-i\varrho\frac{\alpha}{2}}\\ 
\frac{\hbar v(\varsigma k_s-ik_y)}{\mu_s}\sin\varphi e^{-i\varrho\frac{\beta}{2}}\\ 
\end{bmatrix}\\
&\times \exp\big(\varsigma(ik_sx-\varrho\kappa x)+ik_yy\big),
\end{align}
where $k_s=\sqrt{\mu_s^2-(\hbar k_y)^2}/\hbar v$ is the longitudinal wave vector for superconducting region with $\mu_s$ being the Fermi energy of the superconductor, $\kappa=\mu_s\Delta_{sc}\sin\beta/\hbar^2v^2k_s$, $\varsigma=+$ $(-)$ denotes the electron-like (hole-like) quasiparticle states, $\varrho=\pm 1$ and $\varsigma\varrho=+$ $(-)$ for the right (left) propagating states. The phase parameter is given by $\beta=-i\mathrm{arccosh}(E/\Delta_{sc})$ for $E>\Delta_{sc}$ and $\beta=\arccos(E/\Delta_{sc})$ for $E<\Delta_{sc}$.

The probability current $\bm j$ can be obtained from the continuity equation $\partial_t|\Psi|^2+\bm\nabla\cdot\bm j=0$ with the quasiparticle wave function $\Psi=(\psi_A,\psi_B,\psi_C,\phi_A,\phi_B,\phi_C)^T$ satisfying the DBdG Eq.~(\ref{eq:dbdg}), which is given by

\begin{align}
\nonumber j_x=&+v~\mathrm{Re}\big[\psi_B^*(\psi_A\cos\varphi+\psi_C\sin\varphi)\big]\\
&-v~\mathrm{Re}\big[\phi_B^*(\phi_A\cos\varphi+\phi_C\sin\varphi)\big],\label{current1}\\
\nonumber j_y=&-v~\mathrm{Im}\big[\psi_B^*(\psi_A\cos\varphi-\psi_C\sin\varphi)\big]\times\eta\\
&+v~\mathrm{Im}\big[\psi_B^*(\psi_A\cos\varphi-\psi_C\sin\varphi)\big]\times\eta.\label{current2}
\end{align}
The first (second) lines of Eqs.~(\ref{current1}) and (\ref{current2}) come from the contribution of the electron (hole) component of $\Psi$. Consequently, the probability current conservation along $x$ direction requires the continuity of $\psi_B, \phi_B$, $\psi_A\cos\varphi+\psi_C\sin\varphi$, and $\phi_A\cos\varphi+\phi_C\sin\varphi$ at the boundary.

The intervalley electron scattering is neglected due to the large separation of $K$ and $K'$ in momentum space~\cite{PhysRevLett.99.236809,PhysRevLett.98.176805}. As a result, the scattering wave function consists of the intravalley normal reflection/transmission processes and intervalley Andreev reflection/transmission processes, which is given by
\begin{align}
\psi(x)=\begin{cases}
\psi_{\lambda\eta\sigma}^{(e)+}+r\psi_{\lambda\eta\sigma}^{(e)-}+r_A\psi^{(h)-}_{\lambda'\bar{\eta}\bar{\sigma}},& x<0, \\ 
\sum_{\varsigma\varrho}a_{\varsigma\varrho}\psi^{(s)}_{\varsigma\varrho}, &0<x<d, \\ 
t\psi_{\lambda\eta\sigma}^{(e)+}+t_{CAR}\psi^{(h)+}_{\lambda'\bar{\eta}\bar{\sigma}}, &x>d, 
\end{cases}
\end{align}
where $a_{\varsigma\varrho}$ is the scattering amplitude in the superconducting region. In the normal region, $r$, $r_A$, $t$ and $t_{CAR}$ are the scattering amplitudes for the normal reflection, local Andreev reflection, electron elastic cotunneling and crossed Andreev reflection, respectively, which can be obtained by matching the wave functions at the boundary.

\begin{figure}[!tp]
\centerline{\includegraphics[width=0.8\linewidth]{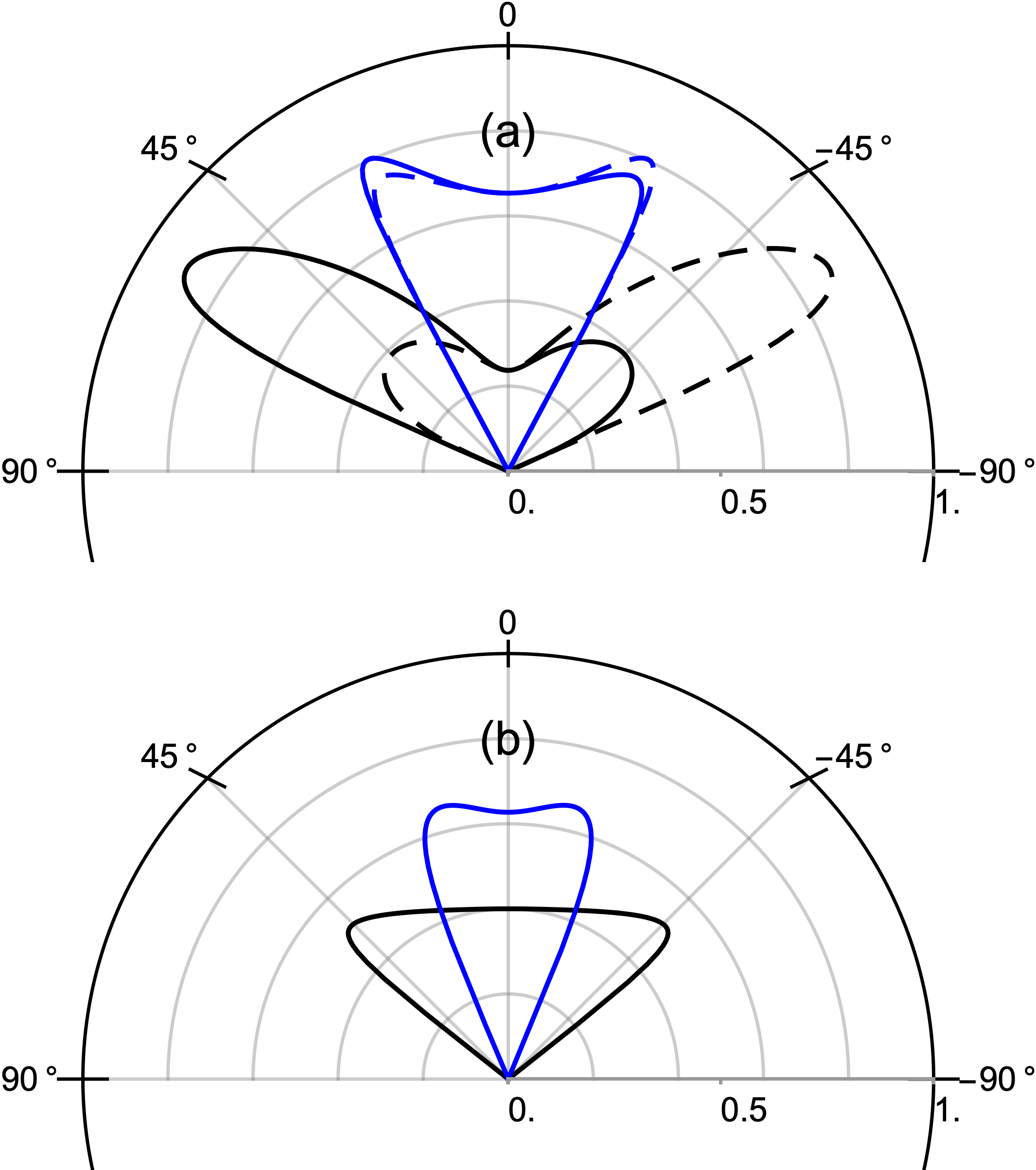}}
\caption{\label{fig:2}
$T_{CAR}$ versus the incident angle $\theta$ for $K$ valley (solid) and $K'$ valley (dashed) at $\mu_s=\SI{120}{\meV}$, $\Delta_{sc}=\SI{1}{\meV}$, $\Delta=\SI{40}{\meV}$, $h=\SI{30}{\meV}$, and $L=1.5\xi_0$ with $\xi_0=\hbar v/\pi\Delta_{sc}$ being the coherence lengths. (a) $T_{CAR}$ for $\alpha=0.4$ (black) and $\alpha=0.8$ (blue).  (b) $T_{CAR}$ for $\alpha=0$ (black) and $\alpha=1$ (blue).
}
\end{figure}

\section{Results}\label{sec:3}

In the numeric calculation, we choose the superconducting pair potential $\Delta_{sc}=\SI{1}{\meV}$, the effective mass $\Delta=\SI{40}{\meV}$, and the exchange splitting energy of the F region is $h=\SI{30}{\meV}$. The pCAR regime appears when the Fermi levels are located between two spin-subband edges, requiring $\SI{10}{meV}<E_{F,1}<\SI{70}{meV}$ and $\SI{-70}{meV}<E_{F,2}<\SI{-10}{meV}$.

The scattering probability of the normal reflection and the CAR can be obtained by $R=|r|^2$ and $T_{CAR}=|j^{h}_x/j^{e}_x||t_{CAR}|^2$, respectively. The conservation of the probability currents requires $R+T_{CAR}=1$. The normal reflection occurs in the same spin subband, leading to the reflection angle $\theta'$ being identical to the incident angle $\theta$, which is given by $\theta'=\theta=\sin^{-1}(\hbar vk_y/(\sqrt{(E+E_{F,1}-\sigma h)^2-\Delta^2)})$ with $\theta\in[-\pi/2,\pi/2]$. For the crossed Andreev reflection, the transmission angle is given by 

\begin{align}
\theta_{CAR}=\sin^{-1}\left[\sqrt{\frac{(E+E_{F,1}-\sigma h)^2-\Delta^2}{(E-E_{F,2}-\sigma h)^2-\Delta^2}}\sin\theta\right].
\end{align}

The angle-resolved CAR probability for $\alpha=0.4$ is shown in Fig.~\ref{fig:2}(a) (black lines). The electron in $K$ valley have a large CAR probability for the incident angles in the range of $45^\circ$ to $60^\circ$. The CARs in $K'$ valley are similarly asymmetric but skewed into the opposite direction. The carries in different valleys turn into different transverse directions, leading to a transverse valley current. This similar valley-contrasting CAR also occurs for $\alpha=0.8$, as shown in Fig.~\ref{fig:2}(a) (blue lines). In fact, for $\alpha\neq0$ and $\alpha\neq1$, this skew CAR for a given valley always exists and the CAR probability is asymmetric:
\begin{gather}
T_{CAR}(\theta,\eta)\neq T_{CAR}(\bar{\theta},\eta).\label{eq:asymmetry}
\end{gather}
Due to the time-reversal symmetry, the CAR probabilities for different valleys satisfy
\begin{gather}
T_{CAR}(\theta,\eta)=T_{CAR}(\bar{\theta}.\bar{\eta}),\label{eq:symmetry}
\end{gather}
Eq.~(\ref{eq:symmetry}) implies that the total CAR is mirror symmetric:
\begin{align}
\nonumber&T_{CAR}(\theta,+)+T_{CAR}(\theta,-)\\=&T_{CAR}(\bar{\theta},+)+T_{CAR}(\bar{\theta},-),
\end{align}
which is responsible for the zero transverse charge current. The CAR probability for $\alpha=0$ and $\alpha=1$ is shown in Fig.~\ref{fig:2}(b), where the valley-contrasting skew CAR is absent.

\begin{figure}[tp]
\centerline{\includegraphics[width=1\linewidth]{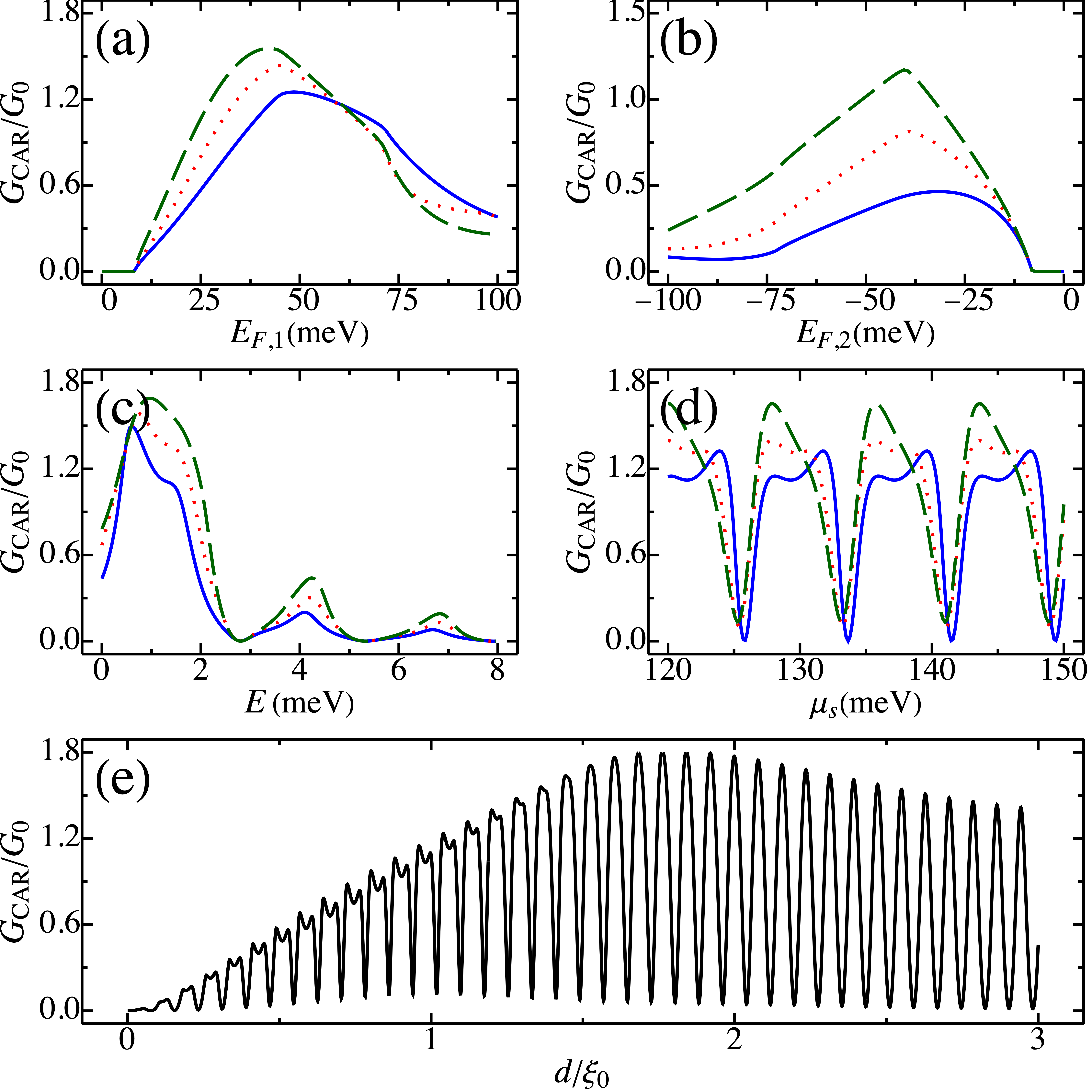}}
\caption{\label{fig:3}
Zero-temperature $G_{CAR}$. (a). $G_{CAR}$ versus $E_{F,1}$ for $\alpha=0$ (solid), $\alpha=0.5$ (dotted), and $\alpha=1$ (dashed) at $E=\SI{1.2}{\meV}$, $E_{F,2}=\SI{-45}{\meV}$, $\mu_{s}=\SI{120}{\meV}$, $\Delta_{sc}=\SI{1}{\meV}$, and $L=1.5\xi_0$. (b) $G_{CAR}$ versus $E_{F,2}$ with $E_{F,1}=\SI{40}{\meV}$ (c) $G_{CAR}$ versus the incident energy $E$ with $(E_{F,1},E_{F,2})=(40,-45)~\mathrm{meV}$. (d) and (e) $G_{CAR}$ versus $\mu_s$ and $d$ with the same Fermi levels in (c).
}
\end{figure}

With the help of the Blonder-Tinkham-Klapwijk (BTK) formula~\cite{PhysRevB.25.4515}, the CAR determined zero-temperature conductance is given by 
\begin{align}
G_{CAR}=\frac{e^2}{2\pi \hbar}\sum_{\eta\sigma}N_\sigma(E)\int d\theta~T_{CAR}(\theta,\eta)\cos\theta,\label{eq:lc}
\end{align}
where $N_\sigma(E)=W\sqrt{(E+E_{F,1}-\sigma h)^2-\Delta^2}/2\pi\hbar v$ is the number of transverse modes for the spin-$\sigma$ channel with $W$ being the junction width. The ballistic conductance of the junction is $G_0=(e^2/2\pi\hbar)\sum_\sigma N_\sigma(E)$. The CAR conductance can be electrically controlled by tuning the incident energy ($E$) and the Fermi level $(E_{F,1(2)},\mu_s)$, and can also be modified by changing the length of the superconducting region ($d$), as shown in Fig.~\ref{fig:3}. $G_{CAR}$ disappears for the bandgap regime $E_{F,1}<(10-E)~\mathrm{meV}=\SI{8.8}{\meV}$ and sharply increases with increasing $E$, as shown in Fig.~\ref{fig:3}(a). For different $\alpha$, $G_{CAR}$ only differs in their amplitudes but exhibits the same increasing tendencies. Tuning $E_{F,2}$ results in the similar characteristics, as shown in Fig.~\ref{fig:3}(b), where the bandgap regime is $E_{F,2}>(-10+E)~\mathrm{meV}=\SI{-8.8}{\meV}$. $G_{CAR}$ versus $E$ is shown in Fig.~\ref{fig:3}(c), the smooth oscillation of $G_{CAR}$ occurs due to the resonant transport in the superconducting region. $G_{CAR}$ vanishes at the resonant energy $E\simeq(2.8,5.2,7.6)~\mathrm{meV}$, where the pCAR disappears for all incident angles. Tuning $\mu_s$ directly modifies the superconducting wave vector $k_s$. The resonant factor $k_s\cdot d$ of the superconducting region leads to the oscillations in Fig.~\ref{fig:3}(d). The $d$-dependence of $G_{CAR}$ is shown in Fig.~\ref{fig:3}(e), the rapid oscillation occurs due to the large superconducting chemical potential $\mu_s$ in the superconducting region. The oscillation peaks appear at $d=n\pi/k_s$ with $n\in\mathbb{Z}$.

In the pCAR regime, the transverse charge current is only carried by the electron states in the left normal region due to the absence of the local Andreev reflected holes, leading to the transverse charge current density for valley $\eta$ \cite{PhysRevLett.115.056602,PhysRevB.100.060507}
\begin{align}
\nonumber J_T^{\eta}=&J_{T}^{\rightarrow}+J_{T}^{\leftarrow}\\
=\nonumber&e\sum_{\sigma}\sum_{\bm k}\left[\frac{\hbar v^2k_y}{E+E_{F,1}-\sigma h}-\frac{\hbar v^2k_y\left|r(\theta,\eta)\right|^2}{E+E_{F,1}-\sigma h}\right]\mathcal{P}^{\rightarrow}_e\\
+&e\sum_{\sigma}\sum_{\bm k}\left[\frac{\hbar v^2k_y}{E-E_{F,1}-\sigma h}-\frac{\hbar v^2k_y\left|\bar{r}(\theta,\eta)\right|^2}{E-E_{F,1}-\sigma h}\right]\mathcal{P}^{\leftarrow}_e,\label{eq:cc}
\end{align}   
where $J_{T}^{\rightarrow}$ and $J_{T}^{\leftarrow}$ denote the transverse current carried by the right and left moving scattering states, respectively, $r$ and $\bar{r}$ are the normal reflection amplitudes corresponding to the incident electron from the left and right, respectively, $\mathcal{P}^{\rightarrow}_e=f_0(E-eV)[1-f_0(E)]$ ($\mathcal{P}^{\leftarrow}_e=[1-f_0(E-eV)]f_0(E)$) indicates that the electron state is occupied in the left (right) and unoccupied in the right (left), where $f_0(E)=1/(e^{E/k_BT}+1)$ is the Fermi distribution function with $k_B$ and $T$ being the Boltzmann constant and temperature, respectively. With the help of the identities: $\sum_{k_y}\frac{k_y}{k_x}|r|^2=-\sum_{k_y}\frac{k_y}{\bar{k}_x}|\bar{r}|^2$, $\int dk_x=\int dE(\partial k_x/\partial E)=\int dE(E+E_{F,1}-\sigma h)/\hbar^2 v^2k_x$ and $df_0(E)/dE=-\delta(E)$ at $T=\SI{0}{\K}$, the zero-temperature transverse charge conductance $\partial J^\eta_T/\partial(eV)$ is given by
\begin{align}
G_T^\eta&=\frac{e^2W}{\hbar}\sum_{\sigma}\int dk_y\frac{k_y}{k_x}\left(1-\left|r(\theta,\eta)\right|^2\right),\\
&=\frac{e^2}{2\pi\hbar}\sum_{\sigma}N_{\sigma}(E)\int d\theta~T_{CAR}(\theta,\eta)\sin\theta.
\end{align} 
Due to the skew CAR given by Eq.~(\ref{eq:asymmetry}), both the $K$ and $K'$ valley generate a nonzero transverse current. With the help of Eqs.~(\ref{eq:asymmetry}) and~(\ref{eq:symmetry}), one finds that $G_T^{K}=-G_{T}^{K'}$, resulting in the transverse valley conductance $G_T^v=G_T^K-G_T^{K'}=2G_{T}^K$ and the transverse charge conductance $G_T=G_T^K+G_T^{K'}=0$. The efficiency of the charge-valley conversion is characterized by the valley Hall angle $\theta_{VH}$, which is given by 
\begin{align}
\tan(\theta_{VH})=\frac{\displaystyle\int d\theta~T_{CAR}(\theta,+)\sin\theta}{\displaystyle\int d\theta~T_{CAR}(\theta,+)\cos\theta}.
\end{align}
\begin{figure}[tp]
\centerline{\includegraphics[width=1\linewidth]{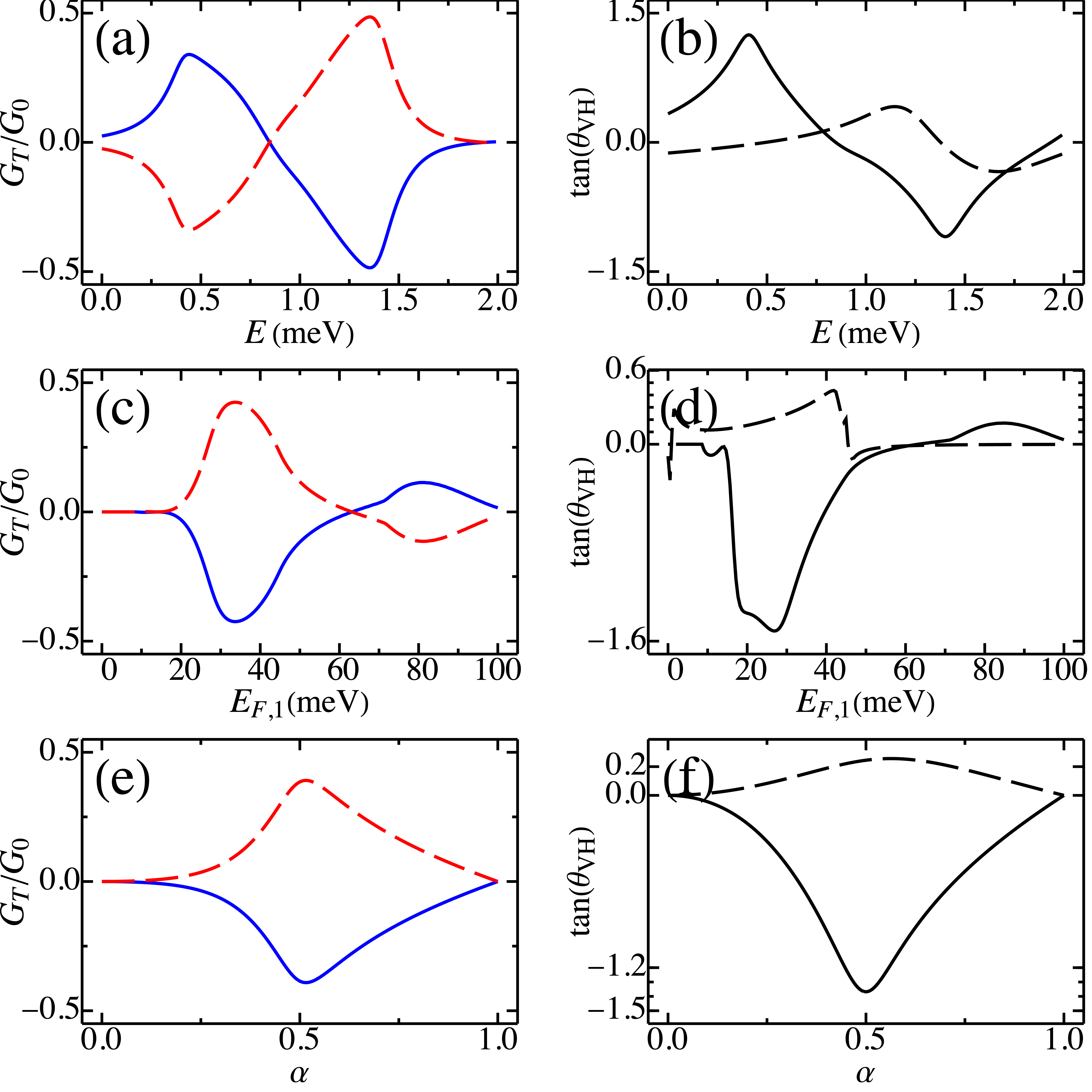}}
\caption{\label{fig:3x}
(a), (c), and (e) Transverse charge current for $K$ valley (solid) and $K'$ valley (dashed). (b), (d), and (f) Tangent of the valley Hall angle for the pCAR regime in $\alpha-\mathcal{T}_3$ based FSF junctions (solid) and the pristine $\alpha-\mathcal{T}_3$ based NSN junctions (dashed). The parameters are same as Fig.~\ref{fig:2}.
}
\end{figure}
$G_{T}^K$ and $G_{T}^{K'}$ exhibit a symmetric pattern when tuning the incident energy as shown in Fig.~\ref{fig:3x}(a), which is attributed to the valley-contrasting CAR. $G_{T}^K$ approaches the maximum value at $E\simeq\SI{0.48}{\meV}$ with the maximum value of the valley Hall angle $\tan(\theta_{VH})\simeq1.4$, as shown in Fig.~\ref{fig:3x}(b). $G_{T}$ and $\tan(\theta_{VH})$ versus the Fermi level $E_{F,1}$ are shown in Figs.~\ref{fig:3x}(c) and~\ref{fig:3x}(d), respectively. The absolute value of the valley Hall angle exhibits a sharply increasing in the pCAR regime, indicating the high efficiency of the charge-valley conversion. The transverse valley current approaches the maximum value at $\alpha\simeq0.5$ with the absolute value of the valley Hall angle $|\tan(\theta_{VH})|\simeq1.5$, as shown in Figs.~\ref{fig:3x}(e). The transverse valley current completely vanishes at $\alpha=0$ and $\alpha=1$ due to the absence of the skew CAR, as shown in Figs.~\ref{fig:3x}(f). It shows that the charge-valley conversion in the CAR assisted model is more efficient than that in the impurity-scattering model ($\tan(\theta_{VH})\simeq0.8$)~\cite{PhysRevB.96.045412}. We note that the valley-contrasting skew CAR also occurs when we go beyond the pCAR regime, where the nonlocal transverse charge and valley conductance are no longer simply determined by the CAR processes. The local Andreev reflection and the electron elastic cotunneling also play an important role. The nonlocal valley Hall angle for the pristine $\alpha-\mathcal{T}_3$ lattices based normal metal/superconductor/normal metal (NSN) junctions ($\Delta=h=0$) is shown in Figs.~\ref{fig:3x}(b),~\ref{fig:3x}(d) and~\ref{fig:3x}(f) with the dashed lines. The maximum absolute value of the valley Hall angle is given by $|\tan(\theta_{VH})|\simeq0.5$, which is generally smaller than that in the pCAR regime. It is believed that the CAR assisted transverse valley current has the similar origin as the skew scattering of the valley-free impurities in $\alpha-\mathcal{T}_3$ lattices, which is attributed to the singular non-$\pi$ Berry curvature caused geometric valley Hall effect~\cite{PhysRevB.96.045412}.

\section{Conclusions}\label{sec:4}

To conclude, we study the nonlocal transport in the ferromagnet-superconductor-ferromagnet junctions base on the $\alpha-\mathcal{T}_3$ lattices, where the spin subbands can be gapped by introducing the symmetry breaking mass term, leading to the pCAR dominated regime. The scattering amplitudes are obtained by solving the DBdG equation, the valley-contrasting skew CAR occurs for $\alpha\neq0$ and $1$, leading to the transverse valley current with zero net charge. The valley Hall angle in the pCAR assisted skew scattering process can reach an order of unity, indicating the high efficiency of the charge-valley conversions.

\section*{Acknowledgements}
This work is supported by the National Key R\&D Program of China (Grant No.\ 2017YFA0303203) and by the NSFC (Grant No.\ 11474149).


\begin{thebibliography}{51}%
\makeatletter
\providecommand \@ifxundefined [1]{%
 \@ifx{#1\undefined}
}%
\providecommand \@ifnum [1]{%
 \ifnum #1\expandafter \@firstoftwo
 \else \expandafter \@secondoftwo
 \fi
}%
\providecommand \@ifx [1]{%
 \ifx #1\expandafter \@firstoftwo
 \else \expandafter \@secondoftwo
 \fi
}%
\providecommand \natexlab [1]{#1}%
\providecommand \enquote  [1]{``#1''}%
\providecommand \bibnamefont  [1]{#1}%
\providecommand \bibfnamefont [1]{#1}%
\providecommand \citenamefont [1]{#1}%
\providecommand \href@noop [0]{\@secondoftwo}%
\providecommand \href [0]{\begingroup \@sanitize@url \@href}%
\providecommand \@href[1]{\@@startlink{#1}\@@href}%
\providecommand \@@href[1]{\endgroup#1\@@endlink}%
\providecommand \@sanitize@url [0]{\catcode `\\12\catcode `\$12\catcode
  `\&12\catcode `\#12\catcode `\^12\catcode `\_12\catcode `\%12\relax}%
\providecommand \@@startlink[1]{}%
\providecommand \@@endlink[0]{}%
\providecommand \url  [0]{\begingroup\@sanitize@url \@url }%
\providecommand \@url [1]{\endgroup\@href {#1}{\urlprefix }}%
\providecommand \urlprefix  [0]{URL }%
\providecommand \Eprint [0]{\href }%
\providecommand \doibase [0]{https://doi.org/}%
\providecommand \selectlanguage [0]{\@gobble}%
\providecommand \bibinfo  [0]{\@secondoftwo}%
\providecommand \bibfield  [0]{\@secondoftwo}%
\providecommand \translation [1]{[#1]}%
\providecommand \BibitemOpen [0]{}%
\providecommand \bibitemStop [0]{}%
\providecommand \bibitemNoStop [0]{.\EOS\space}%
\providecommand \EOS [0]{\spacefactor3000\relax}%
\providecommand \BibitemShut  [1]{\csname bibitem#1\endcsname}%
\let\auto@bib@innerbib\@empty
%</preamble>
\bibitem [{\citenamefont {Chen}\ \emph {et~al.}(2019)\citenamefont {Chen},
  \citenamefont {Xu}, \citenamefont {Wang}, \citenamefont {Liu},\ and\
  \citenamefont {Ma}}]{PhysRevB.99.045420}%
  \BibitemOpen
  \bibfield  {author} {\bibinfo {author} {\bibfnamefont {Y.-R.}\ \bibnamefont
  {Chen}}, \bibinfo {author} {\bibfnamefont {Y.}~\bibnamefont {Xu}}, \bibinfo
  {author} {\bibfnamefont {J.}~\bibnamefont {Wang}}, \bibinfo {author}
  {\bibfnamefont {J.-F.}\ \bibnamefont {Liu}},\ and\ \bibinfo {author}
  {\bibfnamefont {Z.}~\bibnamefont {Ma}},\ }\href
  {https://doi.org/10.1103/PhysRevB.99.045420} {\bibfield  {journal} {\bibinfo
  {journal} {Phys. Rev. B}\ }\textbf {\bibinfo {volume} {99}},\ \bibinfo
  {pages} {045420} (\bibinfo {year} {2019})}\BibitemShut {NoStop}%
\bibitem [{\citenamefont {Dey}\ and\ \citenamefont
  {Ghosh}(2019)}]{PhysRevB.99.205429}%
  \BibitemOpen
  \bibfield  {author} {\bibinfo {author} {\bibfnamefont {B.}~\bibnamefont
  {Dey}}\ and\ \bibinfo {author} {\bibfnamefont {T.~K.}\ \bibnamefont
  {Ghosh}},\ }\href {https://doi.org/10.1103/PhysRevB.99.205429} {\bibfield
  {journal} {\bibinfo  {journal} {Phys. Rev. B}\ }\textbf {\bibinfo {volume}
  {99}},\ \bibinfo {pages} {205429} (\bibinfo {year} {2019})}\BibitemShut
  {NoStop}%
\bibitem [{\citenamefont {Illes}\ and\ \citenamefont
  {Nicol}(2017)}]{PhysRevB.95.235432}%
  \BibitemOpen
  \bibfield  {author} {\bibinfo {author} {\bibfnamefont {E.}~\bibnamefont
  {Illes}}\ and\ \bibinfo {author} {\bibfnamefont {E.~J.}\ \bibnamefont
  {Nicol}},\ }\href {https://doi.org/10.1103/PhysRevB.95.235432} {\bibfield
  {journal} {\bibinfo  {journal} {Phys. Rev. B}\ }\textbf {\bibinfo {volume}
  {95}},\ \bibinfo {pages} {235432} (\bibinfo {year} {2017})}\BibitemShut
  {NoStop}%
\bibitem [{\citenamefont {Weekes}\ \emph {et~al.}(2021)\citenamefont {Weekes},
  \citenamefont {Iurov}, \citenamefont {Zhemchuzhna}, \citenamefont {Gumbs},\
  and\ \citenamefont {Huang}}]{PhysRevB.103.165429}%
  \BibitemOpen
  \bibfield  {author} {\bibinfo {author} {\bibfnamefont {N.}~\bibnamefont
  {Weekes}}, \bibinfo {author} {\bibfnamefont {A.}~\bibnamefont {Iurov}},
  \bibinfo {author} {\bibfnamefont {L.}~\bibnamefont {Zhemchuzhna}}, \bibinfo
  {author} {\bibfnamefont {G.}~\bibnamefont {Gumbs}},\ and\ \bibinfo {author}
  {\bibfnamefont {D.}~\bibnamefont {Huang}},\ }\href
  {https://doi.org/10.1103/PhysRevB.103.165429} {\bibfield  {journal} {\bibinfo
   {journal} {Phys. Rev. B}\ }\textbf {\bibinfo {volume} {103}},\ \bibinfo
  {pages} {165429} (\bibinfo {year} {2021})}\BibitemShut {NoStop}%
\bibitem [{\citenamefont {Dey}\ and\ \citenamefont
  {Ghosh}(2018)}]{PhysRevB.98.075422}%
  \BibitemOpen
  \bibfield  {author} {\bibinfo {author} {\bibfnamefont {B.}~\bibnamefont
  {Dey}}\ and\ \bibinfo {author} {\bibfnamefont {T.~K.}\ \bibnamefont
  {Ghosh}},\ }\href {https://doi.org/10.1103/PhysRevB.98.075422} {\bibfield
  {journal} {\bibinfo  {journal} {Phys. Rev. B}\ }\textbf {\bibinfo {volume}
  {98}},\ \bibinfo {pages} {075422} (\bibinfo {year} {2018})}\BibitemShut
  {NoStop}%
\bibitem [{\citenamefont {Zhu}\ \emph {et~al.}(2017)\citenamefont {Zhu},
  \citenamefont {Zhang}, \citenamefont {Yan}, \citenamefont {Xing},\ and\
  \citenamefont {Zhu}}]{PhysRevA.96.033634}%
  \BibitemOpen
  \bibfield  {author} {\bibinfo {author} {\bibfnamefont {Y.-Q.}\ \bibnamefont
  {Zhu}}, \bibinfo {author} {\bibfnamefont {D.-W.}\ \bibnamefont {Zhang}},
  \bibinfo {author} {\bibfnamefont {H.}~\bibnamefont {Yan}}, \bibinfo {author}
  {\bibfnamefont {D.-Y.}\ \bibnamefont {Xing}},\ and\ \bibinfo {author}
  {\bibfnamefont {S.-L.}\ \bibnamefont {Zhu}},\ }\href
  {https://doi.org/10.1103/PhysRevA.96.033634} {\bibfield  {journal} {\bibinfo
  {journal} {Phys. Rev. A}\ }\textbf {\bibinfo {volume} {96}},\ \bibinfo
  {pages} {033634} (\bibinfo {year} {2017})}\BibitemShut {NoStop}%
\bibitem [{\citenamefont {Nandy}\ \emph {et~al.}(2019)\citenamefont {Nandy},
  \citenamefont {Sengupta},\ and\ \citenamefont {Sen}}]{PhysRevB.100.085134}%
  \BibitemOpen
  \bibfield  {author} {\bibinfo {author} {\bibfnamefont {S.}~\bibnamefont
  {Nandy}}, \bibinfo {author} {\bibfnamefont {K.}~\bibnamefont {Sengupta}},\
  and\ \bibinfo {author} {\bibfnamefont {D.}~\bibnamefont {Sen}},\ }\href
  {https://doi.org/10.1103/PhysRevB.100.085134} {\bibfield  {journal} {\bibinfo
   {journal} {Phys. Rev. B}\ }\textbf {\bibinfo {volume} {100}},\ \bibinfo
  {pages} {085134} (\bibinfo {year} {2019})}\BibitemShut {NoStop}%
\bibitem [{\citenamefont {Cunha}\ \emph {et~al.}(2022)\citenamefont {Cunha},
  \citenamefont {da~Costa}, \citenamefont {Pereira}, \citenamefont {Filho},
  \citenamefont {Van~Duppen},\ and\ \citenamefont
  {Peeters}}]{PhysRevB.105.165402}%
  \BibitemOpen
  \bibfield  {author} {\bibinfo {author} {\bibfnamefont {S.~M.}\ \bibnamefont
  {Cunha}}, \bibinfo {author} {\bibfnamefont {D.~R.}\ \bibnamefont {da~Costa}},
  \bibinfo {author} {\bibfnamefont {J.~M.}\ \bibnamefont {Pereira}}, \bibinfo
  {author} {\bibfnamefont {R.~N.~C.}\ \bibnamefont {Filho}}, \bibinfo {author}
  {\bibfnamefont {B.}~\bibnamefont {Van~Duppen}},\ and\ \bibinfo {author}
  {\bibfnamefont {F.~M.}\ \bibnamefont {Peeters}},\ }\href
  {https://doi.org/10.1103/PhysRevB.105.165402} {\bibfield  {journal} {\bibinfo
   {journal} {Phys. Rev. B}\ }\textbf {\bibinfo {volume} {105}},\ \bibinfo
  {pages} {165402} (\bibinfo {year} {2022})}\BibitemShut {NoStop}%
\bibitem [{\citenamefont {Zeng}\ and\ \citenamefont
  {Shen}(2022{\natexlab{a}})}]{Zeng_2022}%
  \BibitemOpen
  \bibfield  {author} {\bibinfo {author} {\bibfnamefont {W.}~\bibnamefont
  {Zeng}}\ and\ \bibinfo {author} {\bibfnamefont {R.}~\bibnamefont {Shen}},\
  }\href {https://doi.org/10.1088/1367-2630/ac614e} {\bibfield  {journal}
  {\bibinfo  {journal} {New Journal of Physics}\ }\textbf {\bibinfo {volume}
  {24}},\ \bibinfo {pages} {043021} (\bibinfo {year}
  {2022}{\natexlab{a}})}\BibitemShut {NoStop}%
\bibitem [{\citenamefont {Wang}\ and\ \citenamefont
  {Ran}(2011)}]{PhysRevB.84.241103}%
  \BibitemOpen
  \bibfield  {author} {\bibinfo {author} {\bibfnamefont {F.}~\bibnamefont
  {Wang}}\ and\ \bibinfo {author} {\bibfnamefont {Y.}~\bibnamefont {Ran}},\
  }\href {https://doi.org/10.1103/PhysRevB.84.241103} {\bibfield  {journal}
  {\bibinfo  {journal} {Phys. Rev. B}\ }\textbf {\bibinfo {volume} {84}},\
  \bibinfo {pages} {241103} (\bibinfo {year} {2011})}\BibitemShut {NoStop}%
\bibitem [{\citenamefont {Malcolm}\ and\ \citenamefont
  {Nicol}(2015)}]{PhysRevB.92.035118}%
  \BibitemOpen
  \bibfield  {author} {\bibinfo {author} {\bibfnamefont {J.~D.}\ \bibnamefont
  {Malcolm}}\ and\ \bibinfo {author} {\bibfnamefont {E.~J.}\ \bibnamefont
  {Nicol}},\ }\href {https://doi.org/10.1103/PhysRevB.92.035118} {\bibfield
  {journal} {\bibinfo  {journal} {Phys. Rev. B}\ }\textbf {\bibinfo {volume}
  {92}},\ \bibinfo {pages} {035118} (\bibinfo {year} {2015})}\BibitemShut
  {NoStop}%
\bibitem [{\citenamefont {Iurov}\ \emph
  {et~al.}(2020{\natexlab{a}})\citenamefont {Iurov}, \citenamefont
  {Zhemchuzhna}, \citenamefont {Fekete}, \citenamefont {Gumbs},\ and\
  \citenamefont {Huang}}]{PhysRevResearch.2.043245}%
  \BibitemOpen
  \bibfield  {author} {\bibinfo {author} {\bibfnamefont {A.}~\bibnamefont
  {Iurov}}, \bibinfo {author} {\bibfnamefont {L.}~\bibnamefont {Zhemchuzhna}},
  \bibinfo {author} {\bibfnamefont {P.}~\bibnamefont {Fekete}}, \bibinfo
  {author} {\bibfnamefont {G.}~\bibnamefont {Gumbs}},\ and\ \bibinfo {author}
  {\bibfnamefont {D.}~\bibnamefont {Huang}},\ }\href
  {https://doi.org/10.1103/PhysRevResearch.2.043245} {\bibfield  {journal}
  {\bibinfo  {journal} {Phys. Rev. Research}\ }\textbf {\bibinfo {volume}
  {2}},\ \bibinfo {pages} {043245} (\bibinfo {year}
  {2020}{\natexlab{a}})}\BibitemShut {NoStop}%
\bibitem [{\citenamefont {Iurov}\ \emph
  {et~al.}(2020{\natexlab{b}})\citenamefont {Iurov}, \citenamefont
  {Zhemchuzhna}, \citenamefont {Dahal}, \citenamefont {Gumbs},\ and\
  \citenamefont {Huang}}]{PhysRevB.101.035129}%
  \BibitemOpen
  \bibfield  {author} {\bibinfo {author} {\bibfnamefont {A.}~\bibnamefont
  {Iurov}}, \bibinfo {author} {\bibfnamefont {L.}~\bibnamefont {Zhemchuzhna}},
  \bibinfo {author} {\bibfnamefont {D.}~\bibnamefont {Dahal}}, \bibinfo
  {author} {\bibfnamefont {G.}~\bibnamefont {Gumbs}},\ and\ \bibinfo {author}
  {\bibfnamefont {D.}~\bibnamefont {Huang}},\ }\href
  {https://doi.org/10.1103/PhysRevB.101.035129} {\bibfield  {journal} {\bibinfo
   {journal} {Phys. Rev. B}\ }\textbf {\bibinfo {volume} {101}},\ \bibinfo
  {pages} {035129} (\bibinfo {year} {2020}{\natexlab{b}})}\BibitemShut
  {NoStop}%
\bibitem [{\citenamefont {Zhou}(2021)}]{PhysRevB.104.125441}%
  \BibitemOpen
  \bibfield  {author} {\bibinfo {author} {\bibfnamefont {X.}~\bibnamefont
  {Zhou}},\ }\href {https://doi.org/10.1103/PhysRevB.104.125441} {\bibfield
  {journal} {\bibinfo  {journal} {Phys. Rev. B}\ }\textbf {\bibinfo {volume}
  {104}},\ \bibinfo {pages} {125441} (\bibinfo {year} {2021})}\BibitemShut
  {NoStop}%
\bibitem [{\citenamefont {Feng}\ \emph {et~al.}(2020)\citenamefont {Feng},
  \citenamefont {Liu}, \citenamefont {Yu}, \citenamefont {Ma}, \citenamefont
  {Ang}, \citenamefont {Ang},\ and\ \citenamefont
  {Yang}}]{PhysRevB.101.235417}%
  \BibitemOpen
  \bibfield  {author} {\bibinfo {author} {\bibfnamefont {X.}~\bibnamefont
  {Feng}}, \bibinfo {author} {\bibfnamefont {Y.}~\bibnamefont {Liu}}, \bibinfo
  {author} {\bibfnamefont {Z.-M.}\ \bibnamefont {Yu}}, \bibinfo {author}
  {\bibfnamefont {Z.}~\bibnamefont {Ma}}, \bibinfo {author} {\bibfnamefont
  {L.~K.}\ \bibnamefont {Ang}}, \bibinfo {author} {\bibfnamefont {Y.~S.}\
  \bibnamefont {Ang}},\ and\ \bibinfo {author} {\bibfnamefont {S.~A.}\
  \bibnamefont {Yang}},\ }\href {https://doi.org/10.1103/PhysRevB.101.235417}
  {\bibfield  {journal} {\bibinfo  {journal} {Phys. Rev. B}\ }\textbf {\bibinfo
  {volume} {101}},\ \bibinfo {pages} {235417} (\bibinfo {year}
  {2020})}\BibitemShut {NoStop}%
\bibitem [{\citenamefont {Lieb}(1989)}]{PhysRevLett.62.1201}%
  \BibitemOpen
  \bibfield  {author} {\bibinfo {author} {\bibfnamefont {E.~H.}\ \bibnamefont
  {Lieb}},\ }\href {https://doi.org/10.1103/PhysRevLett.62.1201} {\bibfield
  {journal} {\bibinfo  {journal} {Phys. Rev. Lett.}\ }\textbf {\bibinfo
  {volume} {62}},\ \bibinfo {pages} {1201} (\bibinfo {year}
  {1989})}\BibitemShut {NoStop}%
\bibitem [{\citenamefont {Tasaki}(1992)}]{PhysRevLett.69.1608}%
  \BibitemOpen
  \bibfield  {author} {\bibinfo {author} {\bibfnamefont {H.}~\bibnamefont
  {Tasaki}},\ }\href {https://doi.org/10.1103/PhysRevLett.69.1608} {\bibfield
  {journal} {\bibinfo  {journal} {Phys. Rev. Lett.}\ }\textbf {\bibinfo
  {volume} {69}},\ \bibinfo {pages} {1608} (\bibinfo {year}
  {1992})}\BibitemShut {NoStop}%
\bibitem [{\citenamefont {Bodyfelt}\ \emph {et~al.}(2014)\citenamefont
  {Bodyfelt}, \citenamefont {Leykam}, \citenamefont {Danieli}, \citenamefont
  {Yu},\ and\ \citenamefont {Flach}}]{PhysRevLett.113.236403}%
  \BibitemOpen
  \bibfield  {author} {\bibinfo {author} {\bibfnamefont {J.~D.}\ \bibnamefont
  {Bodyfelt}}, \bibinfo {author} {\bibfnamefont {D.}~\bibnamefont {Leykam}},
  \bibinfo {author} {\bibfnamefont {C.}~\bibnamefont {Danieli}}, \bibinfo
  {author} {\bibfnamefont {X.}~\bibnamefont {Yu}},\ and\ \bibinfo {author}
  {\bibfnamefont {S.}~\bibnamefont {Flach}},\ }\href
  {https://doi.org/10.1103/PhysRevLett.113.236403} {\bibfield  {journal}
  {\bibinfo  {journal} {Phys. Rev. Lett.}\ }\textbf {\bibinfo {volume} {113}},\
  \bibinfo {pages} {236403} (\bibinfo {year} {2014})}\BibitemShut {NoStop}%
\bibitem [{\citenamefont {Chalker}\ \emph {et~al.}(2010)\citenamefont
  {Chalker}, \citenamefont {Pickles},\ and\ \citenamefont
  {Shukla}}]{PhysRevB.82.104209}%
  \BibitemOpen
  \bibfield  {author} {\bibinfo {author} {\bibfnamefont {J.~T.}\ \bibnamefont
  {Chalker}}, \bibinfo {author} {\bibfnamefont {T.~S.}\ \bibnamefont
  {Pickles}},\ and\ \bibinfo {author} {\bibfnamefont {P.}~\bibnamefont
  {Shukla}},\ }\href {https://doi.org/10.1103/PhysRevB.82.104209} {\bibfield
  {journal} {\bibinfo  {journal} {Phys. Rev. B}\ }\textbf {\bibinfo {volume}
  {82}},\ \bibinfo {pages} {104209} (\bibinfo {year} {2010})}\BibitemShut
  {NoStop}%
\bibitem [{\citenamefont {Ang}\ \emph {et~al.}(2017)\citenamefont {Ang},
  \citenamefont {Yang}, \citenamefont {Zhang}, \citenamefont {Ma},\ and\
  \citenamefont {Ang}}]{PhysRevB.96.245410}%
  \BibitemOpen
  \bibfield  {author} {\bibinfo {author} {\bibfnamefont {Y.~S.}\ \bibnamefont
  {Ang}}, \bibinfo {author} {\bibfnamefont {S.~A.}\ \bibnamefont {Yang}},
  \bibinfo {author} {\bibfnamefont {C.}~\bibnamefont {Zhang}}, \bibinfo
  {author} {\bibfnamefont {Z.}~\bibnamefont {Ma}},\ and\ \bibinfo {author}
  {\bibfnamefont {L.~K.}\ \bibnamefont {Ang}},\ }\href
  {https://doi.org/10.1103/PhysRevB.96.245410} {\bibfield  {journal} {\bibinfo
  {journal} {Phys. Rev. B}\ }\textbf {\bibinfo {volume} {96}},\ \bibinfo
  {pages} {245410} (\bibinfo {year} {2017})}\BibitemShut {NoStop}%
\bibitem [{\citenamefont {Luo}\ \emph {et~al.}(2020)\citenamefont {Luo},
  \citenamefont {Peng}, \citenamefont {Qu},\ and\ \citenamefont
  {Zhong}}]{PhysRevB.101.245416}%
  \BibitemOpen
  \bibfield  {author} {\bibinfo {author} {\bibfnamefont {C.}~\bibnamefont
  {Luo}}, \bibinfo {author} {\bibfnamefont {X.}~\bibnamefont {Peng}}, \bibinfo
  {author} {\bibfnamefont {J.}~\bibnamefont {Qu}},\ and\ \bibinfo {author}
  {\bibfnamefont {J.}~\bibnamefont {Zhong}},\ }\href
  {https://doi.org/10.1103/PhysRevB.101.245416} {\bibfield  {journal} {\bibinfo
   {journal} {Phys. Rev. B}\ }\textbf {\bibinfo {volume} {101}},\ \bibinfo
  {pages} {245416} (\bibinfo {year} {2020})}\BibitemShut {NoStop}%
\bibitem [{\citenamefont {Schomerus}(2010)}]{PhysRevB.82.165409}%
  \BibitemOpen
  \bibfield  {author} {\bibinfo {author} {\bibfnamefont {H.}~\bibnamefont
  {Schomerus}},\ }\href {https://doi.org/10.1103/PhysRevB.82.165409} {\bibfield
   {journal} {\bibinfo  {journal} {Phys. Rev. B}\ }\textbf {\bibinfo {volume}
  {82}},\ \bibinfo {pages} {165409} (\bibinfo {year} {2010})}\BibitemShut
  {NoStop}%
\bibitem [{\citenamefont {Fripp}\ and\ \citenamefont
  {Kruglyak}(2021)}]{PhysRevB.103.184403}%
  \BibitemOpen
  \bibfield  {author} {\bibinfo {author} {\bibfnamefont {K.~G.}\ \bibnamefont
  {Fripp}}\ and\ \bibinfo {author} {\bibfnamefont {V.~V.}\ \bibnamefont
  {Kruglyak}},\ }\href {https://doi.org/10.1103/PhysRevB.103.184403} {\bibfield
   {journal} {\bibinfo  {journal} {Phys. Rev. B}\ }\textbf {\bibinfo {volume}
  {103}},\ \bibinfo {pages} {184403} (\bibinfo {year} {2021})}\BibitemShut
  {NoStop}%
\bibitem [{\citenamefont {Ezawa}(2013)}]{PhysRevB.87.155415}%
  \BibitemOpen
  \bibfield  {author} {\bibinfo {author} {\bibfnamefont {M.}~\bibnamefont
  {Ezawa}},\ }\href {https://doi.org/10.1103/PhysRevB.87.155415} {\bibfield
  {journal} {\bibinfo  {journal} {Phys. Rev. B}\ }\textbf {\bibinfo {volume}
  {87}},\ \bibinfo {pages} {155415} (\bibinfo {year} {2013})}\BibitemShut
  {NoStop}%
\bibitem [{\citenamefont {Zeng}\ and\ \citenamefont
  {Shen}(2022{\natexlab{b}})}]{PhysRevB.105.094510}%
  \BibitemOpen
  \bibfield  {author} {\bibinfo {author} {\bibfnamefont {W.}~\bibnamefont
  {Zeng}}\ and\ \bibinfo {author} {\bibfnamefont {R.}~\bibnamefont {Shen}},\
  }\href {https://doi.org/10.1103/PhysRevB.105.094510} {\bibfield  {journal}
  {\bibinfo  {journal} {Phys. Rev. B}\ }\textbf {\bibinfo {volume} {105}},\
  \bibinfo {pages} {094510} (\bibinfo {year} {2022}{\natexlab{b}})}\BibitemShut
  {NoStop}%
\bibitem [{\citenamefont {Bouhadida}\ \emph {et~al.}(2020)\citenamefont
  {Bouhadida}, \citenamefont {Mandhour},\ and\ \citenamefont
  {Charfi-Kaddour}}]{PhysRevB.102.075443}%
  \BibitemOpen
  \bibfield  {author} {\bibinfo {author} {\bibfnamefont {F.}~\bibnamefont
  {Bouhadida}}, \bibinfo {author} {\bibfnamefont {L.}~\bibnamefont
  {Mandhour}},\ and\ \bibinfo {author} {\bibfnamefont {S.}~\bibnamefont
  {Charfi-Kaddour}},\ }\href {https://doi.org/10.1103/PhysRevB.102.075443}
  {\bibfield  {journal} {\bibinfo  {journal} {Phys. Rev. B}\ }\textbf {\bibinfo
  {volume} {102}},\ \bibinfo {pages} {075443} (\bibinfo {year}
  {2020})}\BibitemShut {NoStop}%
\bibitem [{\citenamefont {Islam}\ and\ \citenamefont
  {Dutta}(2017)}]{PhysRevB.96.045418}%
  \BibitemOpen
  \bibfield  {author} {\bibinfo {author} {\bibfnamefont {S.~F.}\ \bibnamefont
  {Islam}}\ and\ \bibinfo {author} {\bibfnamefont {P.}~\bibnamefont {Dutta}},\
  }\href {https://doi.org/10.1103/PhysRevB.96.045418} {\bibfield  {journal}
  {\bibinfo  {journal} {Phys. Rev. B}\ }\textbf {\bibinfo {volume} {96}},\
  \bibinfo {pages} {045418} (\bibinfo {year} {2017})}\BibitemShut {NoStop}%
\bibitem [{\citenamefont {Filusch}\ \emph {et~al.}(2021)\citenamefont
  {Filusch}, \citenamefont {Bishop}, \citenamefont {Saxena}, \citenamefont
  {Wellein},\ and\ \citenamefont {Fehske}}]{PhysRevB.103.165114}%
  \BibitemOpen
  \bibfield  {author} {\bibinfo {author} {\bibfnamefont {A.}~\bibnamefont
  {Filusch}}, \bibinfo {author} {\bibfnamefont {A.~R.}\ \bibnamefont {Bishop}},
  \bibinfo {author} {\bibfnamefont {A.}~\bibnamefont {Saxena}}, \bibinfo
  {author} {\bibfnamefont {G.}~\bibnamefont {Wellein}},\ and\ \bibinfo {author}
  {\bibfnamefont {H.}~\bibnamefont {Fehske}},\ }\href
  {https://doi.org/10.1103/PhysRevB.103.165114} {\bibfield  {journal} {\bibinfo
   {journal} {Phys. Rev. B}\ }\textbf {\bibinfo {volume} {103}},\ \bibinfo
  {pages} {165114} (\bibinfo {year} {2021})}\BibitemShut {NoStop}%
\bibitem [{\citenamefont {Hirsch}(1999)}]{PhysRevLett.83.1834}%
  \BibitemOpen
  \bibfield  {author} {\bibinfo {author} {\bibfnamefont {J.~E.}\ \bibnamefont
  {Hirsch}},\ }\href {https://doi.org/10.1103/PhysRevLett.83.1834} {\bibfield
  {journal} {\bibinfo  {journal} {Phys. Rev. Lett.}\ }\textbf {\bibinfo
  {volume} {83}},\ \bibinfo {pages} {1834} (\bibinfo {year}
  {1999})}\BibitemShut {NoStop}%
\bibitem [{\citenamefont {Bernevig}\ and\ \citenamefont
  {Zhang}(2006)}]{PhysRevLett.96.106802}%
  \BibitemOpen
  \bibfield  {author} {\bibinfo {author} {\bibfnamefont {B.~A.}\ \bibnamefont
  {Bernevig}}\ and\ \bibinfo {author} {\bibfnamefont {S.-C.}\ \bibnamefont
  {Zhang}},\ }\href {https://doi.org/10.1103/PhysRevLett.96.106802} {\bibfield
  {journal} {\bibinfo  {journal} {Phys. Rev. Lett.}\ }\textbf {\bibinfo
  {volume} {96}},\ \bibinfo {pages} {106802} (\bibinfo {year}
  {2006})}\BibitemShut {NoStop}%
\bibitem [{\citenamefont {Sinova}\ \emph {et~al.}(2004)\citenamefont {Sinova},
  \citenamefont {Culcer}, \citenamefont {Niu}, \citenamefont {Sinitsyn},
  \citenamefont {Jungwirth},\ and\ \citenamefont
  {MacDonald}}]{PhysRevLett.92.126603}%
  \BibitemOpen
  \bibfield  {author} {\bibinfo {author} {\bibfnamefont {J.}~\bibnamefont
  {Sinova}}, \bibinfo {author} {\bibfnamefont {D.}~\bibnamefont {Culcer}},
  \bibinfo {author} {\bibfnamefont {Q.}~\bibnamefont {Niu}}, \bibinfo {author}
  {\bibfnamefont {N.~A.}\ \bibnamefont {Sinitsyn}}, \bibinfo {author}
  {\bibfnamefont {T.}~\bibnamefont {Jungwirth}},\ and\ \bibinfo {author}
  {\bibfnamefont {A.~H.}\ \bibnamefont {MacDonald}},\ }\href
  {https://doi.org/10.1103/PhysRevLett.92.126603} {\bibfield  {journal}
  {\bibinfo  {journal} {Phys. Rev. Lett.}\ }\textbf {\bibinfo {volume} {92}},\
  \bibinfo {pages} {126603} (\bibinfo {year} {2004})}\BibitemShut {NoStop}%
\bibitem [{\citenamefont {Xiao}\ \emph {et~al.}(2007)\citenamefont {Xiao},
  \citenamefont {Yao},\ and\ \citenamefont {Niu}}]{PhysRevLett.99.236809}%
  \BibitemOpen
  \bibfield  {author} {\bibinfo {author} {\bibfnamefont {D.}~\bibnamefont
  {Xiao}}, \bibinfo {author} {\bibfnamefont {W.}~\bibnamefont {Yao}},\ and\
  \bibinfo {author} {\bibfnamefont {Q.}~\bibnamefont {Niu}},\ }\href
  {https://doi.org/10.1103/PhysRevLett.99.236809} {\bibfield  {journal}
  {\bibinfo  {journal} {Phys. Rev. Lett.}\ }\textbf {\bibinfo {volume} {99}},\
  \bibinfo {pages} {236809} (\bibinfo {year} {2007})}\BibitemShut {NoStop}%
\bibitem [{\citenamefont {Tong}\ \emph {et~al.}(2016)\citenamefont {Tong},
  \citenamefont {Gong}, \citenamefont {Wan},\ and\ \citenamefont
  {Duan}}]{tong2016concepts}%
  \BibitemOpen
  \bibfield  {author} {\bibinfo {author} {\bibfnamefont {W.-Y.}\ \bibnamefont
  {Tong}}, \bibinfo {author} {\bibfnamefont {S.-J.}\ \bibnamefont {Gong}},
  \bibinfo {author} {\bibfnamefont {X.}~\bibnamefont {Wan}},\ and\ \bibinfo
  {author} {\bibfnamefont {C.-G.}\ \bibnamefont {Duan}},\ }\href
  {https://www.nature.com/articles/ncomms13612} {\bibfield  {journal} {\bibinfo
   {journal} {Nature communications}\ }\textbf {\bibinfo {volume} {7}},\
  \bibinfo {pages} {1} (\bibinfo {year} {2016})}\BibitemShut {NoStop}%
\bibitem [{\citenamefont {Lee}\ \emph {et~al.}(2016)\citenamefont {Lee},
  \citenamefont {Mak},\ and\ \citenamefont {Shan}}]{lee2016electrical}%
  \BibitemOpen
  \bibfield  {author} {\bibinfo {author} {\bibfnamefont {J.}~\bibnamefont
  {Lee}}, \bibinfo {author} {\bibfnamefont {K.~F.}\ \bibnamefont {Mak}},\ and\
  \bibinfo {author} {\bibfnamefont {J.}~\bibnamefont {Shan}},\ }\href
  {https://www.science.org/doi/abs/10.1126/science.1250140?casa_token=Eiz4KlDMpqEAAAAA:OxGeYyIl02aJU9d0Ez2h98Lk6SbZaQ16dM6sVIT1jqf6t1Uln6VnxHugVzCrg8nP4csrC1qMC7OpWEs}
  {\bibfield  {journal} {\bibinfo  {journal} {Nature nanotechnology}\ }\textbf
  {\bibinfo {volume} {11}},\ \bibinfo {pages} {421} (\bibinfo {year}
  {2016})}\BibitemShut {NoStop}%
\bibitem [{\citenamefont {Xu}\ \emph {et~al.}(2017)\citenamefont {Xu},
  \citenamefont {Huang}, \citenamefont {Huang},\ and\ \citenamefont
  {Lai}}]{PhysRevB.96.045412}%
  \BibitemOpen
  \bibfield  {author} {\bibinfo {author} {\bibfnamefont {H.-Y.}\ \bibnamefont
  {Xu}}, \bibinfo {author} {\bibfnamefont {L.}~\bibnamefont {Huang}}, \bibinfo
  {author} {\bibfnamefont {D.}~\bibnamefont {Huang}},\ and\ \bibinfo {author}
  {\bibfnamefont {Y.-C.}\ \bibnamefont {Lai}},\ }\href
  {https://doi.org/10.1103/PhysRevB.96.045412} {\bibfield  {journal} {\bibinfo
  {journal} {Phys. Rev. B}\ }\textbf {\bibinfo {volume} {96}},\ \bibinfo
  {pages} {045412} (\bibinfo {year} {2017})}\BibitemShut {NoStop}%
\bibitem [{\citenamefont {Beckmann}\ \emph {et~al.}(2004)\citenamefont
  {Beckmann}, \citenamefont {Weber},\ and\ \citenamefont
  {v.~L\"ohneysen}}]{PhysRevLett.93.197003}%
  \BibitemOpen
  \bibfield  {author} {\bibinfo {author} {\bibfnamefont {D.}~\bibnamefont
  {Beckmann}}, \bibinfo {author} {\bibfnamefont {H.~B.}\ \bibnamefont
  {Weber}},\ and\ \bibinfo {author} {\bibfnamefont {H.}~\bibnamefont
  {v.~L\"ohneysen}},\ }\href {https://doi.org/10.1103/PhysRevLett.93.197003}
  {\bibfield  {journal} {\bibinfo  {journal} {Phys. Rev. Lett.}\ }\textbf
  {\bibinfo {volume} {93}},\ \bibinfo {pages} {197003} (\bibinfo {year}
  {2004})}\BibitemShut {NoStop}%
\bibitem [{\citenamefont {Benjamin}\ and\ \citenamefont
  {Pachos}(2008)}]{PhysRevB.78.235403}%
  \BibitemOpen
  \bibfield  {author} {\bibinfo {author} {\bibfnamefont {C.}~\bibnamefont
  {Benjamin}}\ and\ \bibinfo {author} {\bibfnamefont {J.~K.}\ \bibnamefont
  {Pachos}},\ }\href {https://doi.org/10.1103/PhysRevB.78.235403} {\bibfield
  {journal} {\bibinfo  {journal} {Phys. Rev. B}\ }\textbf {\bibinfo {volume}
  {78}},\ \bibinfo {pages} {235403} (\bibinfo {year} {2008})}\BibitemShut
  {NoStop}%
\bibitem [{\citenamefont {Law}\ \emph {et~al.}(2009)\citenamefont {Law},
  \citenamefont {Lee},\ and\ \citenamefont {Ng}}]{PhysRevLett.103.237001}%
  \BibitemOpen
  \bibfield  {author} {\bibinfo {author} {\bibfnamefont {K.~T.}\ \bibnamefont
  {Law}}, \bibinfo {author} {\bibfnamefont {P.~A.}\ \bibnamefont {Lee}},\ and\
  \bibinfo {author} {\bibfnamefont {T.~K.}\ \bibnamefont {Ng}},\ }\href
  {https://doi.org/10.1103/PhysRevLett.103.237001} {\bibfield  {journal}
  {\bibinfo  {journal} {Phys. Rev. Lett.}\ }\textbf {\bibinfo {volume} {103}},\
  \bibinfo {pages} {237001} (\bibinfo {year} {2009})}\BibitemShut {NoStop}%
\bibitem [{\citenamefont {Kalenkov}\ and\ \citenamefont
  {Zaikin}(2007)}]{PhysRevB.76.224506}%
  \BibitemOpen
  \bibfield  {author} {\bibinfo {author} {\bibfnamefont {M.~S.}\ \bibnamefont
  {Kalenkov}}\ and\ \bibinfo {author} {\bibfnamefont {A.~D.}\ \bibnamefont
  {Zaikin}},\ }\href {https://doi.org/10.1103/PhysRevB.76.224506} {\bibfield
  {journal} {\bibinfo  {journal} {Phys. Rev. B}\ }\textbf {\bibinfo {volume}
  {76}},\ \bibinfo {pages} {224506} (\bibinfo {year} {2007})}\BibitemShut
  {NoStop}%
\bibitem [{\citenamefont {Jakobsen}\ \emph {et~al.}(2021)\citenamefont
  {Jakobsen}, \citenamefont {Brataas},\ and\ \citenamefont
  {Qaiumzadeh}}]{PhysRevLett.127.017701}%
  \BibitemOpen
  \bibfield  {author} {\bibinfo {author} {\bibfnamefont {M.~F.}\ \bibnamefont
  {Jakobsen}}, \bibinfo {author} {\bibfnamefont {A.}~\bibnamefont {Brataas}},\
  and\ \bibinfo {author} {\bibfnamefont {A.}~\bibnamefont {Qaiumzadeh}},\
  }\href {https://doi.org/10.1103/PhysRevLett.127.017701} {\bibfield  {journal}
  {\bibinfo  {journal} {Phys. Rev. Lett.}\ }\textbf {\bibinfo {volume} {127}},\
  \bibinfo {pages} {017701} (\bibinfo {year} {2021})}\BibitemShut {NoStop}%
\bibitem [{\citenamefont {Zhang}\ and\ \citenamefont
  {Trauzettel}(2019)}]{PhysRevLett.122.257701}%
  \BibitemOpen
  \bibfield  {author} {\bibinfo {author} {\bibfnamefont {S.-B.}\ \bibnamefont
  {Zhang}}\ and\ \bibinfo {author} {\bibfnamefont {B.}~\bibnamefont
  {Trauzettel}},\ }\href {https://doi.org/10.1103/PhysRevLett.122.257701}
  {\bibfield  {journal} {\bibinfo  {journal} {Phys. Rev. Lett.}\ }\textbf
  {\bibinfo {volume} {122}},\ \bibinfo {pages} {257701} (\bibinfo {year}
  {2019})}\BibitemShut {NoStop}%
\bibitem [{\citenamefont {Zhang}\ \emph {et~al.}(2021)\citenamefont {Zhang},
  \citenamefont {Li}, \citenamefont {Pe\~na Benitez}, \citenamefont
  {Sur\'owka}, \citenamefont {Moessner}, \citenamefont {Molenkamp},\ and\
  \citenamefont {Trauzettel}}]{PhysRevLett.127.076601}%
  \BibitemOpen
  \bibfield  {author} {\bibinfo {author} {\bibfnamefont {S.-B.}\ \bibnamefont
  {Zhang}}, \bibinfo {author} {\bibfnamefont {C.-A.}\ \bibnamefont {Li}},
  \bibinfo {author} {\bibfnamefont {F.}~\bibnamefont {Pe\~na Benitez}},
  \bibinfo {author} {\bibfnamefont {P.}~\bibnamefont {Sur\'owka}}, \bibinfo
  {author} {\bibfnamefont {R.}~\bibnamefont {Moessner}}, \bibinfo {author}
  {\bibfnamefont {L.~W.}\ \bibnamefont {Molenkamp}},\ and\ \bibinfo {author}
  {\bibfnamefont {B.}~\bibnamefont {Trauzettel}},\ }\href
  {https://doi.org/10.1103/PhysRevLett.127.076601} {\bibfield  {journal}
  {\bibinfo  {journal} {Phys. Rev. Lett.}\ }\textbf {\bibinfo {volume} {127}},\
  \bibinfo {pages} {076601} (\bibinfo {year} {2021})}\BibitemShut {NoStop}%
\bibitem [{\citenamefont {Zhang}\ \emph {et~al.}(2017)\citenamefont {Zhang},
  \citenamefont {Hou}, \citenamefont {Xie},\ and\ \citenamefont
  {Sun}}]{PhysRevB.95.245433}%
  \BibitemOpen
  \bibfield  {author} {\bibinfo {author} {\bibfnamefont {Y.-T.}\ \bibnamefont
  {Zhang}}, \bibinfo {author} {\bibfnamefont {Z.}~\bibnamefont {Hou}}, \bibinfo
  {author} {\bibfnamefont {X.~C.}\ \bibnamefont {Xie}},\ and\ \bibinfo {author}
  {\bibfnamefont {Q.-F.}\ \bibnamefont {Sun}},\ }\href
  {https://doi.org/10.1103/PhysRevB.95.245433} {\bibfield  {journal} {\bibinfo
  {journal} {Phys. Rev. B}\ }\textbf {\bibinfo {volume} {95}},\ \bibinfo
  {pages} {245433} (\bibinfo {year} {2017})}\BibitemShut {NoStop}%
\bibitem [{\citenamefont {Ang}\ \emph {et~al.}(2016)\citenamefont {Ang},
  \citenamefont {Ang}, \citenamefont {Zhang},\ and\ \citenamefont
  {Ma}}]{PhysRevB.93.041422}%
  \BibitemOpen
  \bibfield  {author} {\bibinfo {author} {\bibfnamefont {Y.~S.}\ \bibnamefont
  {Ang}}, \bibinfo {author} {\bibfnamefont {L.~K.}\ \bibnamefont {Ang}},
  \bibinfo {author} {\bibfnamefont {C.}~\bibnamefont {Zhang}},\ and\ \bibinfo
  {author} {\bibfnamefont {Z.}~\bibnamefont {Ma}},\ }\href
  {https://doi.org/10.1103/PhysRevB.93.041422} {\bibfield  {journal} {\bibinfo
  {journal} {Phys. Rev. B}\ }\textbf {\bibinfo {volume} {93}},\ \bibinfo
  {pages} {041422} (\bibinfo {year} {2016})}\BibitemShut {NoStop}%
\bibitem [{\citenamefont {Uchida}\ \emph {et~al.}(2019)\citenamefont {Uchida},
  \citenamefont {Koretsune}, \citenamefont {Sato}, \citenamefont {Kriener},
  \citenamefont {Nakazawa}, \citenamefont {Nishihaya}, \citenamefont {Taguchi},
  \citenamefont {Arita},\ and\ \citenamefont {Kawasaki}}]{PhysRevB.100.245148}%
  \BibitemOpen
  \bibfield  {author} {\bibinfo {author} {\bibfnamefont {M.}~\bibnamefont
  {Uchida}}, \bibinfo {author} {\bibfnamefont {T.}~\bibnamefont {Koretsune}},
  \bibinfo {author} {\bibfnamefont {S.}~\bibnamefont {Sato}}, \bibinfo {author}
  {\bibfnamefont {M.}~\bibnamefont {Kriener}}, \bibinfo {author} {\bibfnamefont
  {Y.}~\bibnamefont {Nakazawa}}, \bibinfo {author} {\bibfnamefont
  {S.}~\bibnamefont {Nishihaya}}, \bibinfo {author} {\bibfnamefont
  {Y.}~\bibnamefont {Taguchi}}, \bibinfo {author} {\bibfnamefont
  {R.}~\bibnamefont {Arita}},\ and\ \bibinfo {author} {\bibfnamefont
  {M.}~\bibnamefont {Kawasaki}},\ }\href
  {https://doi.org/10.1103/PhysRevB.100.245148} {\bibfield  {journal} {\bibinfo
   {journal} {Phys. Rev. B}\ }\textbf {\bibinfo {volume} {100}},\ \bibinfo
  {pages} {245148} (\bibinfo {year} {2019})}\BibitemShut {NoStop}%
\bibitem [{\citenamefont {Beenakker}(2006)}]{PhysRevLett.97.067007}%
  \BibitemOpen
  \bibfield  {author} {\bibinfo {author} {\bibfnamefont {C.~W.~J.}\
  \bibnamefont {Beenakker}},\ }\href
  {https://doi.org/10.1103/PhysRevLett.97.067007} {\bibfield  {journal}
  {\bibinfo  {journal} {Phys. Rev. Lett.}\ }\textbf {\bibinfo {volume} {97}},\
  \bibinfo {pages} {067007} (\bibinfo {year} {2006})}\BibitemShut {NoStop}%
\bibitem [{\citenamefont {Gennes}(2018)}]{de2018superconductivity}%
  \BibitemOpen
  \bibfield  {author} {\bibinfo {author} {\bibfnamefont {P.-G.~D.}\
  \bibnamefont {Gennes}},\ }\href@noop {} {\emph {\bibinfo {title}
  {\textit{Superconductivity of metals and alloys}}}}\ (\bibinfo  {publisher}
  {CRC, Boca Raton, FL},\ \bibinfo {year} {2018})\BibitemShut {NoStop}%
\bibitem [{\citenamefont {Gorbachev}\ \emph {et~al.}(2007)\citenamefont
  {Gorbachev}, \citenamefont {Tikhonenko}, \citenamefont {Mayorov},
  \citenamefont {Horsell},\ and\ \citenamefont
  {Savchenko}}]{PhysRevLett.98.176805}%
  \BibitemOpen
  \bibfield  {author} {\bibinfo {author} {\bibfnamefont {R.~V.}\ \bibnamefont
  {Gorbachev}}, \bibinfo {author} {\bibfnamefont {F.~V.}\ \bibnamefont
  {Tikhonenko}}, \bibinfo {author} {\bibfnamefont {A.~S.}\ \bibnamefont
  {Mayorov}}, \bibinfo {author} {\bibfnamefont {D.~W.}\ \bibnamefont
  {Horsell}},\ and\ \bibinfo {author} {\bibfnamefont {A.~K.}\ \bibnamefont
  {Savchenko}},\ }\href {https://doi.org/10.1103/PhysRevLett.98.176805}
  {\bibfield  {journal} {\bibinfo  {journal} {Phys. Rev. Lett.}\ }\textbf
  {\bibinfo {volume} {98}},\ \bibinfo {pages} {176805} (\bibinfo {year}
  {2007})}\BibitemShut {NoStop}%
\bibitem [{\citenamefont {Blonder}\ \emph {et~al.}(1982)\citenamefont
  {Blonder}, \citenamefont {Tinkham},\ and\ \citenamefont
  {Klapwijk}}]{PhysRevB.25.4515}%
  \BibitemOpen
  \bibfield  {author} {\bibinfo {author} {\bibfnamefont {G.~E.}\ \bibnamefont
  {Blonder}}, \bibinfo {author} {\bibfnamefont {M.}~\bibnamefont {Tinkham}},\
  and\ \bibinfo {author} {\bibfnamefont {T.~M.}\ \bibnamefont {Klapwijk}},\
  }\href {https://doi.org/10.1103/PhysRevB.25.4515} {\bibfield  {journal}
  {\bibinfo  {journal} {Phys. Rev. B}\ }\textbf {\bibinfo {volume} {25}},\
  \bibinfo {pages} {4515} (\bibinfo {year} {1982})}\BibitemShut {NoStop}%
\bibitem [{\citenamefont {Matos-Abiague}\ and\ \citenamefont
  {Fabian}(2015)}]{PhysRevLett.115.056602}%
  \BibitemOpen
  \bibfield  {author} {\bibinfo {author} {\bibfnamefont {A.}~\bibnamefont
  {Matos-Abiague}}\ and\ \bibinfo {author} {\bibfnamefont {J.}~\bibnamefont
  {Fabian}},\ }\href {https://doi.org/10.1103/PhysRevLett.115.056602}
  {\bibfield  {journal} {\bibinfo  {journal} {Phys. Rev. Lett.}\ }\textbf
  {\bibinfo {volume} {115}},\ \bibinfo {pages} {056602} (\bibinfo {year}
  {2015})}\BibitemShut {NoStop}%
\bibitem [{\citenamefont {Costa}\ \emph {et~al.}(2019)\citenamefont {Costa},
  \citenamefont {Matos-Abiague},\ and\ \citenamefont
  {Fabian}}]{PhysRevB.100.060507}%
  \BibitemOpen
  \bibfield  {author} {\bibinfo {author} {\bibfnamefont {A.}~\bibnamefont
  {Costa}}, \bibinfo {author} {\bibfnamefont {A.}~\bibnamefont
  {Matos-Abiague}},\ and\ \bibinfo {author} {\bibfnamefont {J.}~\bibnamefont
  {Fabian}},\ }\href {https://doi.org/10.1103/PhysRevB.100.060507} {\bibfield
  {journal} {\bibinfo  {journal} {Phys. Rev. B}\ }\textbf {\bibinfo {volume}
  {100}},\ \bibinfo {pages} {060507} (\bibinfo {year} {2019})}\BibitemShut
  {NoStop}%
\end{thebibliography}
\end{document}